\documentclass[aps,prl,twocolumn,superscriptaddress]{revtex4}

\usepackage{graphicx}
\usepackage{dcolumn}
\usepackage{bm}

\begin{document}
\newcommand{\lao}{LaAlO$_3$} 
\newcommand{\sto}{SrTiO$_3$} 
\newcommand{\lsato}{La$_{1-x}$Sr$_x$Al$_{1-y}$Ti$_y$O$_3$} 


\title{Intrinsic origin of the two-dimensional electron gas at polar oxide interfaces}


\author{M.L.~Reinle-Schmitt}
 \affiliation{Paul Scherrer Institut, CH-5232 Villigen, Switzerland}
\author{C.~Cancellieri}
 \affiliation{Paul Scherrer Institut, CH-5232 Villigen, Switzerland}
\author{D.~Li}
 \affiliation{DPMC, University of Geneva, 24 Quai Ernest-Ansermet,
 1211 Gen\`eve 4, Switzerland} 
\author{D. Fontaine} 
 \affiliation{Physique Th\'eorique des Mat\'eriaux, Universit\'e de
 Li\`ege, B-4000 Li\`ege, Belgium} 
\author{M.~Medarde}
 \affiliation{Paul Scherrer Institut, CH-5232 Villigen, Switzerland}
\author{E.~Pomjakushina}
 \affiliation{Paul Scherrer Institut, CH-5232 Villigen, Switzerland}
\author{C.W.~Schneider}
 \affiliation{Paul Scherrer Institut, CH-5232 Villigen, Switzerland}
\author{S. Gariglio}
 \affiliation{DPMC, University of Geneva, 24 Quai Ernest-Ansermet,
 1211 Gen\`eve 4, Switzerland} 
\author{Ph. Ghosez} 
 \affiliation{Physique Th\'eorique des Mat\'eriaux, Universit\'e de
 Li\`ege, B-4000 Li\`ege, Belgium} 
\author{J.-M. Triscone}
 \affiliation{DPMC, University of Geneva, 24 Quai Ernest-Ansermet,
 1211 Gen\`eve 4, Switzerland} 
\author{P.R.~Willmott}
 \email[]{philip.willmott@psi.ch}
 \affiliation{Paul Scherrer Institut, CH-5232 Villigen, Switzerland}

\date{\today}

\begin{abstract}
The predictions of the polar catastrophe scenario to explain the occurrence of a metallic interface in heterostructures of the solid solution(LaAlO$_3$)$_{x}$(SrTiO$_3$)$_{1-x}$ (LASTO:x) grown on (001)~SrTiO$_3$ were investigated as a function of film thickness and $x$. The films are insulating for the thinnest layers, but above a critical thickness, $t_c$, the interface exhibits a constant finite conductivity which depends in a predictable manner on $x$. It is shown that $t_c$ scales with the strength of the built-in electric field of the polar material, and is immediately understandable in terms of an electronic reconstruction at the nonpolar-polar interface. These results thus conclusively identify the polar-catastrophe model as the intrinsic origin of the doping at this polar oxide interface. 
\end{abstract}


\maketitle


Conductivity at the \lao/\sto\ (LAO/STO) interface was originally explained in terms of the so-called polar-catastrophe scenario \cite{Ohtomo2004,Nakagawa2006,Thiel2006}. In this model, an {\em intrinsic} electronic reconstruction at the interface is expected from the buildup of an internal electrical potential in LAO as the film thickness $t$ increases, due to the polar discontinuity at the interface between LAO, which consists of alternating positively and negatively charged layers, (LaO)$^{+}$ and (AlO$_2$)$^{-}$, and STO, with charge-neutral layers. However, models explaining the conductivity in terms of {\em extrinsic} effects caused by structural deviations from a perfect interface have also been proposed \cite{Huijben2006,Siemons2007,Herranz2007,Kalabukhov2007,Willmott2007,Kalabukhov2009,Qiao2010,Qiao2011,Bristowe2011}.

\begin{figure*}
\includegraphics[scale=0.31,angle=-90]{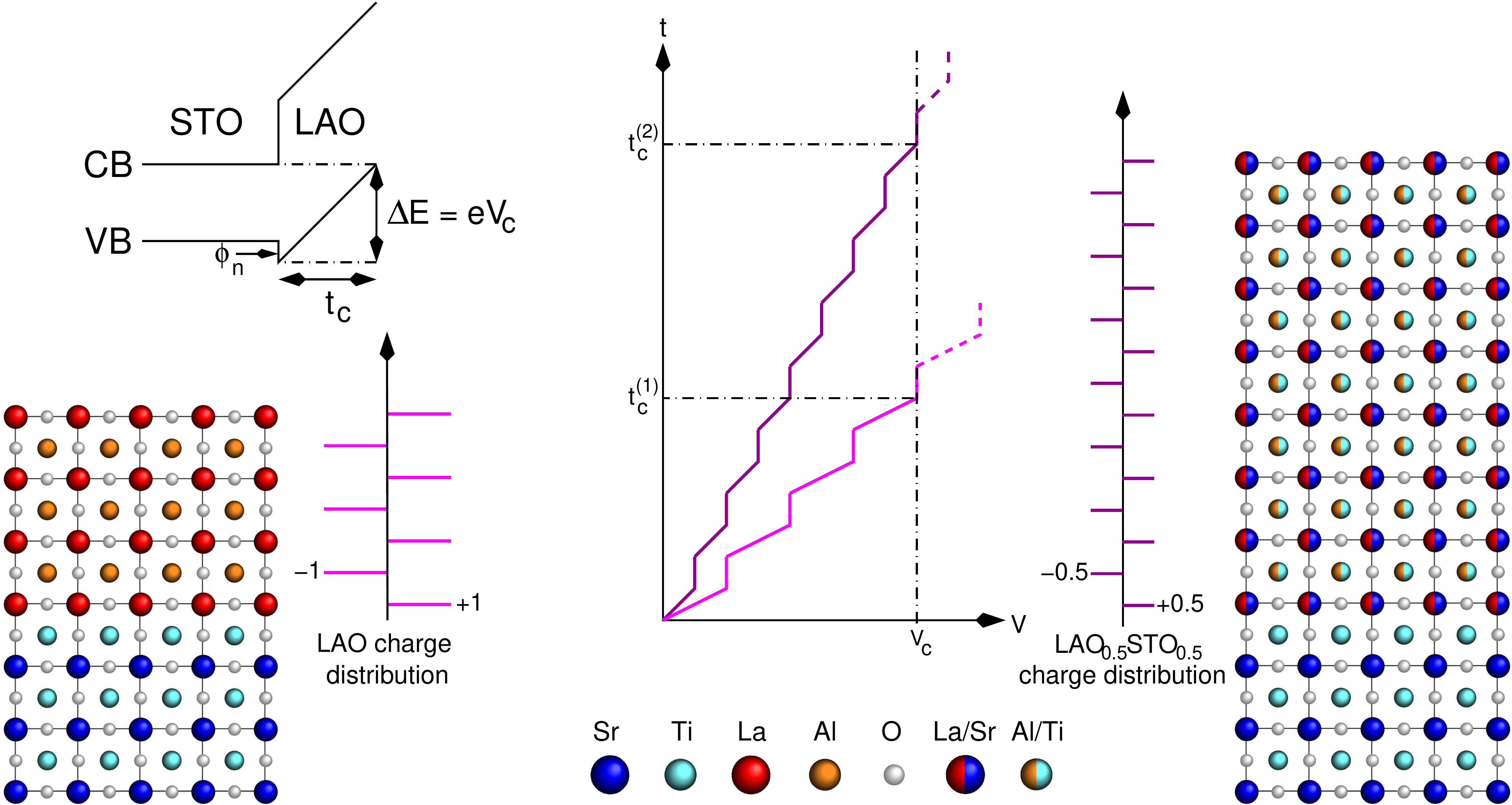}
\caption{(color online) Schematic showing the buildup of potential in a polar layer as a function of its thickness $t$ for LAO and LASTO:$0.5$, assuming the same relative permittivity but different formal polarization $P_0$ induced by the charge of the successive A-site and B-site sublayers. The critical thicknesses for the electronic reconstruction are labeled $t_c^{(1)}$ and $t_c^{(2)}$ for LAO and LASTO:$0.5$, respectively. The band-level scheme shows band bending in the pure LAO layer of the valence band (VB) and conduction band (CB), and the critical thickness $t_c$ and potential buildup $eV_c$ required to induce the electronic reconstruction. $\phi_n$ is the valence-band offset between STO and LAO.}
\label{fig1}
\end{figure*}

The intrinsic doping mechanism, illustrated in Fig.~\ref{fig1}, was elegantly reformulated in the framework of the modern theory of polarization \cite{Stengel2009}: LAO has a {\it formal} polarization $P^0_{\rm LAO}= e/2S= 0.529$~C\,m$^{-2}$ (where $S$ is the unit-cell cross-section in the plane of the interface), while in nonpolar STO, $P^0_{\rm STO}= 0$. The preservation of the normal component of the electric displacement field $D$ along the STO/LAO/vacuum stack in the absence of free charge at the surface and interface ($D=0$) requires the appearance of a macroscopic electric field ${\cal E}_{\rm LAO}$. Due to the dielectric response of LAO, this field is ${\cal E}_{\rm LAO} = P_0/\varepsilon_0 \varepsilon_{\rm LAO}=0.24$~V\,\AA$^{-1}$, where $\varepsilon_{\rm LAO} \approx 24$ is the relative permittivity of LAO \cite{Konaka1991}. 

The electric field in LAO will bend the electronic bands as illustrated in Fig.~\ref{fig1}. At a thickness $t_c$, the valence O~$2p$ bands of LAO at the surface reach the level of the STO Ti~$3d$ conduction bands at the interface and a Zener breakdown occurs. Above this thickness, electrons will be transferred progressively from the surface to the interface, which hence becomes metallic. This simple electrostatic model not only explains the conduction, but also links the formal polarization of LAO, its dielectric constant, and $t_c$ such that 
\begin{eqnarray} 
\label{eq:tc} 
t_c = \frac{\varepsilon_0 \varepsilon_{\rm LAO} \Delta E}{e P^0_{\rm LAO}}, 
\end{eqnarray} 
where $\Delta E \approx 3.3$ eV is the difference of energy between the valence band of LAO and the conduction band of STO and $e$ is the electron charge. This yields an estimate of $t_c \approx 3.5$~monolayers (MLs).  

The main success of the polar-catastrophe scenario is that it very
accurately predicts the critical thickness $t_c$ originally observed
experimentally by Thiel~{\em et al.} \cite{Thiel2006,Schlom2011natmat}, a very robust result which has been replicated in several laboratories \cite{Huijben2006,Cancellieri2010} and also predicted from first-principles calculations on ideal STO/LAO/vacuum stacks \cite{Popovic2008,Lee2008}. However, the surface of LAO and its interface with STO are far from ideal and alternative, extrinsic, effects have been invoked to explain the observed conductivity, such as oxygen vacancies \cite{Siemons2007,Herranz2007,Kalabukhov2007}, adsorbates at the LAO surface \cite{Bristowe2011}, or intermixing between LAO and STO at the interface \cite{Willmott2007,Kalabukhov2009,Qiao2010,Qiao2011}. The presence of an electric field within the LAO layer, expected to produce the Zener breakdown, was also questioned \cite{Segal2009}, although recent surface x-ray diffraction measurements revealed an atomic rumpling in the LAO layer, a clear signature of an electric field \cite{Pauli2011}, as well as an expansion of the LAO $c$-axis in very thin layers, compatible with an electrostrictive effect produced by ${\cal E}_{\rm LAO}$ \cite{Cancellieri2011}. 

In this Letter we describe experiments performed to further assess the relative importance of the intrinsic polar-catastrophe scenario and the extrinsic intermixing model to explain the origin of the conductivity. To achieve this, we replaced pure LAO with ultrathin films of LAO diluted with STO [(LaAlO$_3$)$_{x}$(SrTiO$_3$)$_{1-x}$, or LASTO:$x$] for different values of $x$. This system is both intermixed and has a formal polarization different to that of pure LAO. In this manner, we could observe whether $t_c$ properly evolves with the formal polarization and dielectric constant of this material, as predicted by Eq.~(\ref{eq:tc}), while also investigating the possible role of intermixing. 

Varying the composition of the LASTO:$x$ films (i.e., $x$) allows one to tune continuously the formal polarization such that $P^0_{{\rm LASTO:}x} = x P^0_{\rm LAO}$. For instance, for $x=0.5$, the random solid solution has alternating planes with $+0.5$ and $-0.5$~formal charges, compared to $+1$ and $-1$ charges in pure LAO (see Fig.~\ref{fig1}). One must, however, also consider possible changes of the other fundamental quantities determining the critical thickness expressed in Eq.~(\ref{eq:tc}). The energy gap $\Delta E$ formally depends on the electronic bandgap of STO and the valence-band offset $\phi_n$ (see Fig.~\ref{fig1}) between the two materials, which can evolve with $x$. In practice, however, the O~$2p$ valence bands of STO and LAO ($x=1$) are virtually aligned ($\phi_n = 0.1 $ eV \cite{Popovic2008}) and $\phi_n$ further diminishes with $x$, so that we can confidently approximate $\Delta E \approx E^g_{\rm STO}$, irrespective of the composition. The dielectric constants of STO and LAO, however, differ significantly ($\varepsilon_{\rm STO}=300, \varepsilon_{\rm LAO}=24$ at room temperature) and it is not obvious {\em a~priori} how the dielectric constant of the solid solution will evolve with composition. To clarify this point, we performed first-principles calculations on bulk compounds of different compositions using a supercell technique (see \cite{supplementaryMaterials}). Although the dielectric constant of the LASTO:$x$ evolves slightly with the atomic arrangement, we observe that it remains essentially constant for $x= 1$, $0.75$ and $0.5$. This result may initially seem surprising; however, the large dielectric constant of pure STO is mainly produced by a low-frequency and highly polar phonon mode related to its incipient ferroelectric character. This mode, absent in LAO, is very sensitive to atomic disorder and is stabilized at much larger frequencies through mixing with other modes for $x>0.5$ without contributing significantly to the dielectric constant. Hence, from the discussion above, it appears that varying the composition of the LASTO:$x$ films allows one to tune the formal polarization while keeping the other quantities in Eq.~(\ref{eq:tc}) essentially constant, that is, 
\begin{eqnarray} 
t_c^{{\rm LASTO:}x}=t_c^{\rm LAO}/x. 
\end{eqnarray} 
Based on this argument, we therefore predict that $t_c=7$~ML and $5$~ML for $x=0.5$ and $0.75$, respectively, a result confirmed for $x=0.50$ from first-principles calculations (see \cite{supplementaryMaterials}). 


Two independent series of LASTO:$x$ films were prepared by pulsed laser deposition (PLD) using two different sets of growth parameters. The first set of films, produced at the Paul Scherrer Institut, was grown on both native and TiO$_2$-terminated (001) STO substrates using two different PLD sintered targets, with $x = 0.50$ and $0.75$. Neither target ($> 85$~\% dense) was conducting. Growth conditions using $266$-nm Nd:YAG laser radiation were: pulse energy $= 16$~mJ ($\approx 2$~J\,cm$^{-2}$), $10$~Hz; $T = 750$~$^{\circ}$C, $p_{\rm O_2} = 2.5 \times 10^{-8}$~mbar; sample cooled after growth at $25^{\circ}$C\, min$^{-1}$ and postannealed for one hour in $1$~atm. O$_2$ at $550^{\circ}$C. The second set of films was grown on TiO$_2$-terminated STO at the University of Geneva using standard growth conditions: KrF laser ($248$~nm) with a pulse energy of $50$~mJ ($\approx 0.6$~J\,cm$^{-2}$), $1$~Hz; $T = 800$~$^{\circ}$C, $p_{\rm O_2} = 1 \times 10^{-4}$~mbar; sample cooled after growth to $550^{\circ}$C in $200$~mbar O$_2$ and maintained at this temperature and pressure for one hour before being cooled to room temperature in the same atmosphere. The stoichiometries of the films from both sets were shown by Rutherford backscattering (RBS) to be equal to the nominal PLD-target compositions of $0.5$ and $0.75$ to within the experimental accuracy of $1.5$~\%. In-situ reflection high-energy electron-diffraction (RHEED) measurements, and x-ray diffraction measurements confirming perfectly strained growth are detailed in \cite{supplementaryMaterials}. 

\begin{figure}
\includegraphics[scale=0.76,angle=-90]{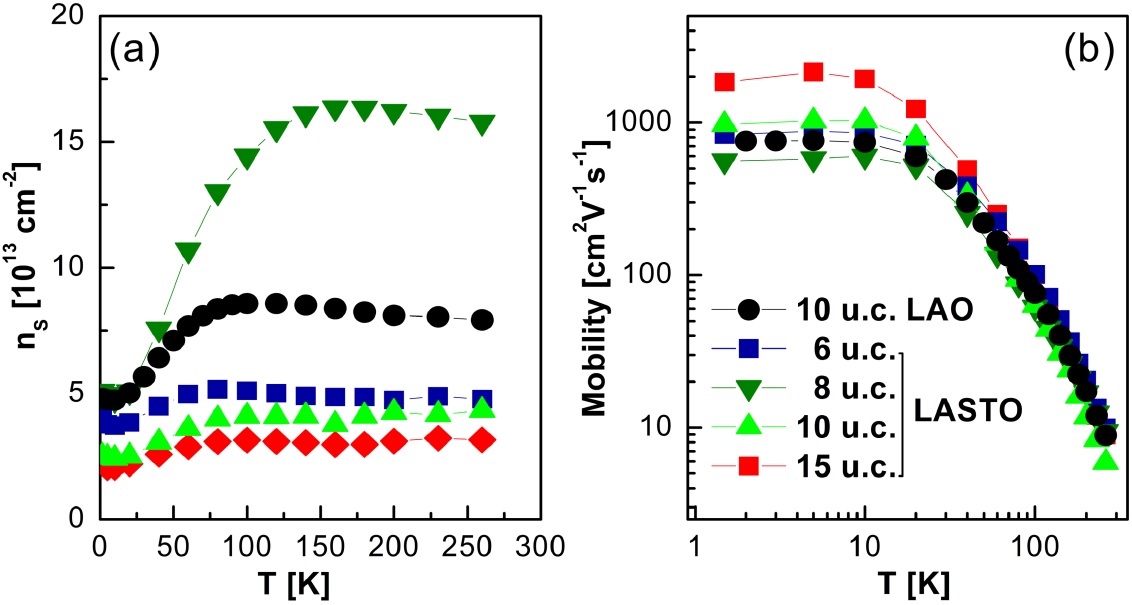}
\caption{(color online) (a) Sheet carrier density and (b) mobility as a function of temperature and film thickness for pure LAO and LASTO:$0.5$.}
\label{fig2}
\end{figure} 

Neither of the mixed-composition film stoichiometries investigated produce layers which increase in conductance with thickness, as one might otherwise expect for intermixed materials which were intrinsically electrically conducting. In addition, none of the films are conducting at the top surface, but instead require careful bonding at the interface to exhibit conductivity. These metallic interfaces were characterized by transport properties using the van der Pauw method. All samples remained metallic down to the lowest measured temperature of $1.5$~K. The sheet carrier densities $n_s$ estimated from the Hall effect at low magnetic field for interfaces with LAO and with LASTO:$0.5$ of different thicknesses are shown in Fig.~\ref{fig2}(a). The value of $n_s$ is in the range $3$ to $15 \times 10^{13}$~cm$^{-2}$, though with no obvious dependence on composition or thickness of the layers. According to the polar-catastrophe model, we should expect LASTO:$x$ samples to exhibit a lower carrier density, as the screening charge scales as $x\cdot e/2S$, whereby $S$ is the unit-cell surface area. However, as already observed for LAO/STO interfaces, the estimation of the carrier density from the Hall effect yields values up to one order of magnitude smaller than those predicted from theory, possibly suggesting a large amount of trapped interface charges. Figure~\ref{fig2}(b) shows the Hall mobility $\mu$ of the same interfaces measured as a function of temperature. Note that interfaces with LAO and LASTO:$x$ display similar values, with $\mu$ larger for samples with low $n_s$.

To probe experimentally the dielectric constant, capacitors were fabricated with different thicknesses of pure LAO and the $x=0.5$ films, using palladium as the top electrode. A serious complication in measurements of such ultrathin films is the significant contribution of the electrode--oxide interface on the capacitance \cite{Stengel2006}, which means the results can only be viewed semi-quantitatively. We observe that the dielectric constant of the LASTO:$0.5$ and LAO display values in the range of $20$ to $30$, and are in good agreement with previous reports on ceramic solid-state solutions, where no large enhancement of the relative permittivity was observed for the solid solution up to $80$~\% STO \cite{Sun1998}. Measurements of the temperature dependence and the electric-field tunability confirm that LASTO:$x$ behaves like LAO rather than STO. Cooling the films to $4$~K produces a small change of the dielectric constant, in sharp contrast with the low-temperature divergence of STO. These experimental results show that there is no large enhancement of the dielectric constant in LASTO:$0.5$ with respect to pure LAO, as predicted by our first-principles calculations.


\begin{figure}
\includegraphics[scale=0.57]{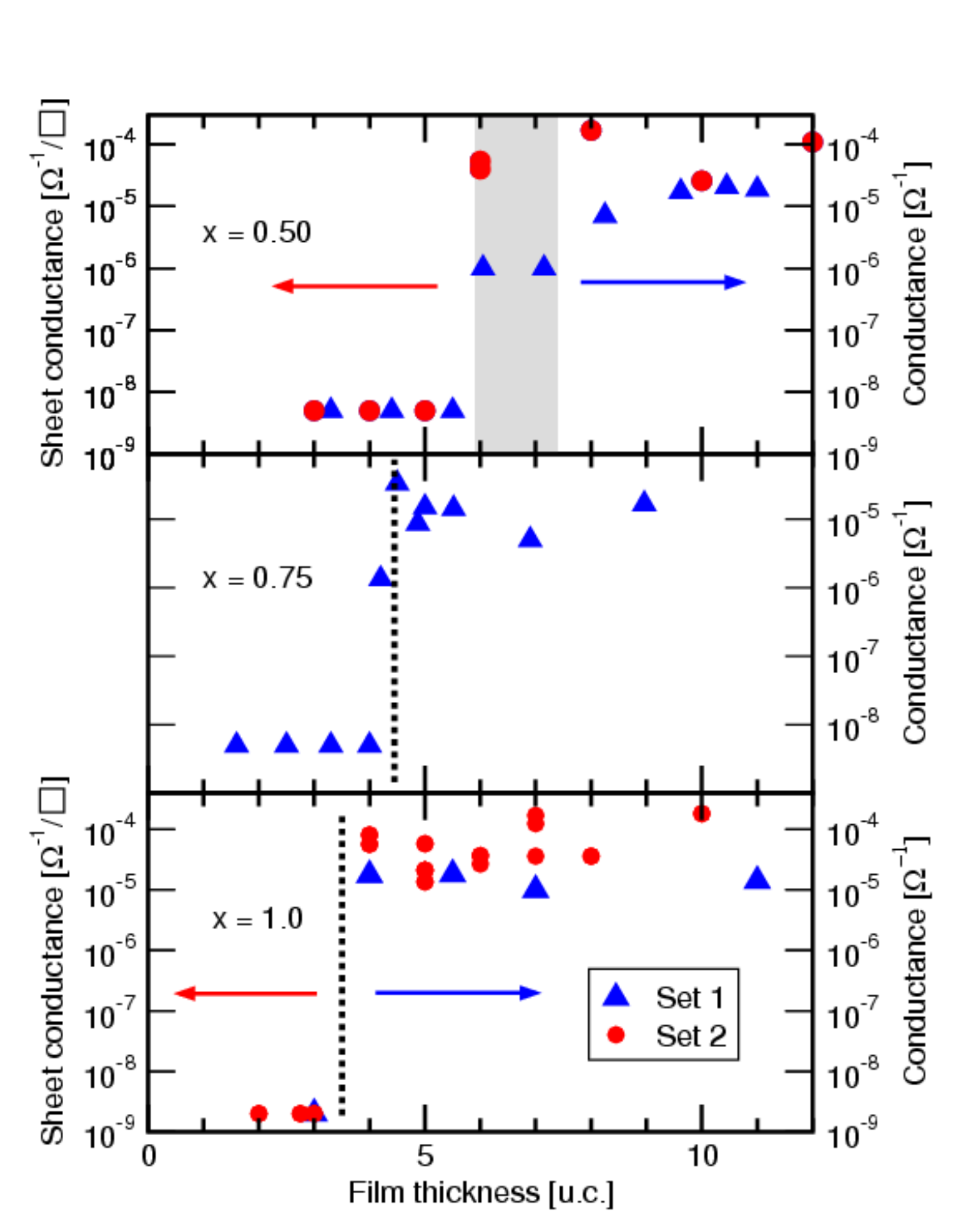}
\caption{\label{fig:conductivity4} (color online) Room-temperature conductance of LASTO:$x$ films for $x = 1$, $0.75$ and $0.50$. The dashed vertical lines for $x=1.0$ and $0.75$ indicate the experimentally determined threshold thicknesses $t_c$, which for $x=0.5$, is represented by a band for the more gradual transition. All values were obtained after ensuring that the samples had remained in dark conditions for a sufficiently long time to avoid any photoelectric contributions.}
\end{figure} 

Figure~\ref{fig:conductivity4} is the central result of this work. The conductance of the interface as a function of the LASTO:$x$ film thickness is shown for $x=1.0$ (pure LAO, lower panel), $x=0.75$ (middle panel), and $x=0.50$ (top panel). The conductivity is given in sheet conductance (left axis) and/or conductance (right axis). The step in conductance for $x=1.0$ is observed at $4$~unit cells, as first observed by Thiel {\em et al.} \cite{Thiel2006} and reproduced by several groups. For $x=0.75$ and $0.50$, the data unambiguously demonstrate that the critical thickness increases with STO-content in the solid solution, with $t_c^{{\rm LASTO:}0.75}$ close to $5$~unit cells, and $t_c^{{\rm LASTO:}0.5}$ between $6$ and $7$~unit cells. This striking result demonstrates that the critical thickness depends on $x$, increasing as the formal polarization decreases. Although the critical thicknesses obtained from the experimental data are marginally smaller than predicted by theory, this can easily be attributed to the uncertainty in the dielectric constants of the solid solutions. The overall agreement, however, with the polar-catastrophe model described by Eq.~(\ref{eq:tc}) is remarkable. 

In conclusion, we have shown that in heterostructures of ultrathin films of the solid solution (LaAlO$_3$)$_x$(SrTiO$_3$)$_{1-x}$ grown on (001)~SrTiO$_3$, the critical thickness at which conductivity is observed scales with the strength of the built-in electric field of the polar material. These results test the fundamental predictions of the polar-catastrophe scenario and convincingly demonstrate the intrinsic origin of the doping at the LAO/STO interface.


\begin{acknowledgments}
Support of this work by the Schweizerischer Nationalfonds zur F\"orderung der wissenschaftlichen Forschung, in particular the National Center of Competence in Research, Materials with Novel Electronic Properties, MaNEP, and by the European Union through the project OxIDes. The staff of the Swiss Light Source is gratefully acknowledged. The authors also thank Dr. Max D\"obeli for the RBS measurements and Dr. Pavlo Zubko for assistance in the capacitance measurements. Ph.G. is grateful to the Francqui Foundation. 
\end{acknowledgments}

\bibliography{bibliography}
\end{document}